\documentclass{article}

\usepackage{PRIMEarxiv}
\usepackage{xcolor}
 \usepackage{amsmath}
\usepackage[utf8]{inputenc} 
\usepackage[T1]{fontenc}    
\usepackage{hyperref}       
\usepackage{url}            
\usepackage{booktabs}       
\usepackage{amsfonts}       
\usepackage{nicefrac}       
\usepackage{microtype}      
\usepackage{cite}
\usepackage{lipsum}
\usepackage{fancyhdr}       
\usepackage{graphicx}       
\graphicspath{{media/}}     

\title{GaN/InN HEMT Based UV Photodetector on SiC with Hexagonal Boron Nitride Passivation
\thanks{\textit{\underline{Citation}}: 
\textbf{Authors. Title. Pages.... DOI:000000/11111.}} 
}

\author{
  Mustafa Kilin \\
  Medical Science and Optics Department\\
  Adiyaman University\\
  \texttt{mkilin@adiyaman.edu.tr} \\
   \And
  Firat Yasar \\
  Boeing Research and Technology\\ Huntington Beach CA 92647\\
  \texttt{firat.yasar@boeing.com} \\
  }

\begin{document}
\maketitle

\begin{abstract}
This work presents a novel Gallium nitride (GaN) High-electron-mobility transistor (HEMT) based ultraviolet photodetector architecture integrating advanced material and structural design strategies to enhance detection performance and stability under room-temperature operation. The device is constructed on a high-thermal-conductivity silicon carbide (SiC) substrate and incorporates an n-GaN buffer, an indium nitride (InN) channel layer for improved electron mobility and two-dimensional electron gas (2DEG) confinement, and a dual-passivation scheme combining silicon nitride (SiN) and hexagonal boron nitride (h-BN). A p-GaN layer is embedded between the passivation interfaces to deplete the 2DEG in dark conditions. Lateral Nickel (Ni) source and drain electrodes and a recessed gate positioned within the substrate ensure enhanced electric field control and noise suppression. Numerical simulations demonstrate that the integration of a hexagonal boron nitride (h-BN) interlayer within the dual passivation stack effectively suppresses the gate leakage current from typical literature values of the order of $10^{-7}$~A to approximately $10^{-10}$~A, highlighting its critical role in enhancing interfacial insulation. In addition, consistent with previous reports, the use of a silicon carbide (SiC) substrate offers significantly improved thermal management over sapphire, enabling more stable operation under UV illumination. The device demonstrates strong photoresponse under 360~nm ultraviolet (UV) illumination, high photo-to-dark current ratios (PDCR) found is approximately $10^{6}$, and tunable performance via structural optimization of p-GaN width between 0.40 $\mu\text{m}$ and 1.60 $\mu\text{m}$ , doping concentration from $5\times10^{16}$ $\text{cm}^{-3}$ to $5\times10^{18}$ $\text{cm}^{-3}$ , and embedding depth between 0.060 $\mu\text{m}$ and 0.068 $\mu\text{m}$ . The results underscore the proposed structure’s notable effectiveness in passivation quality, suppression of gate leakage, and thermal management, collectively establishing it as a robust and reliable platform for next-generation UV photodetectors operating under harsh environmental conditions.
\end{abstract}


\section{Introduction}

Wide-bandgap III-nitride semiconductors have emerged as critical material platforms for UV photodetection technologies intended for deployment in thermally and radiatively harsh environments\cite{Pargam2024}. Their intrinsic solar-blind response, chemical inertness, and mechanical durability have positioned them as attractive candidates for sensing tasks ranging from atmospheric diagnostics to spaceborne surveillance and flame tracking systems \cite{alaie2015uv, luo2022uv, Liu2023Fac}. GaN-based heterostructures, in particular, benefit from high breakdown electric field, strong carrier confinement, and compatibility with advanced epitaxial growth, which enable stable operation at elevated temperatures \cite{luo2023physstatus}. A variety of device topologies—including Schottky barrier photodiodes \cite{ye2024schottky}, metal semiconductor metal (MSM) photodetectors \cite{kilin2025sim,Sayran2025msm}, and p–i–n or avalanche diodes \cite{Blain2025pin, Zhu2025Aval} have been developed using these materials, though their performance often remains constrained by weak internal gain and temperature-sensitive dark current behavior. To overcome these limitations, polar heterostructures such as GaN/InN have gained interest for their ability to support 2DEG formation at the interface via strong polarization fields \cite{armstrong2019polarization, Mondal2024Enh, He2025Rev}. While the 2DEG facilitates enhanced lateral electron transport and responsivity under illumination, it also introduces significant leakage pathways under dark conditions, particularly in continuous biasing regimes. Recent advances in polarization-aware band engineering, gate insulation strategies using high-k dielectrics such as h-BN \cite{He2025Rev}, and magnesium (Mg) doped p-GaN modulation layers \cite{Chiu2025High, Liu2025High} have demonstrated promising results in balancing photocurrent gain and dark-state suppression. Moreover, optimized channel geometries and thermal design—especially with thermally conductive substrates like SiC—have been shown to further stabilize electrical characteristics under high-intensity UV exposure \cite{Dai2015Enh, Pan2025Des}.These collective insights point toward the necessity of co-optimizing gate control, material interfaces, and thermal pathways in next-generation UV photodetector architectures.

To mitigate the adverse effects associated with parasitic 2DEG formation—particularly elevated leakage currents and reduced signal-to-noise ratios—advanced structural strategies have been actively explored~\cite{Talukder2025Comp, Zhang2020Supp}. Among these, embedding gate-modulating p-type regions within the gate stack has shown significant promise in controlling the 2DEG distribution without compromising responsivity~\cite{Chen2025Res}.When appropriately aligned with the channel, the p-GaN layer facilitates charge depletion via built-in electric fields, effectively suppressing dark current while enabling light-triggered recovery of the conductive channel. In parallel, precise geometric engineering of the gate region, including recessed or substrate-buried gate configurations, enhances electric field localization and lateral confinement, yielding greater controllability over photoresponsive transitions \cite{Sreelekshmi2025Rec, Li2025Pro}. Furthermore, gate dielectric innovation has introduced metal–insulator–semiconductor (MIS) architectures as a practical means of minimizing gate leakage and stabilizing interfacial properties under bias stress\cite{Chen2016Tow}. Incorporating high-quality dielectrics such as h-BN has demonstrated notable improvements in leakage suppression and interface trap passivation \cite{Jnug2024Def}. The anisotropic bonding and atomically smooth surface of h-BN provide superior electrical isolation compared to conventional SiN, which often suffers from higher defect densities and thermally induced breakdown \cite{Chai2016Mak}. Moreover, thermomechanical stability of the overall device architecture can be significantly enhanced by substrate selection; in particular, SiC substrates support effective heat dissipation and reduce self-heating-induced degradation, as confirmed in both experimental and modeling studies \cite{Lee2014Anal, Bao2017Imp, Valdemar2021Thermo }. These integrated approaches—spanning electrostatic modulation, gate dielectric refinement, and thermal interface management—collectively establish a multifactorial design framework for high-performance and stable UV photodetectors based on III-nitride heterostructures\cite{Muñoz2001Nitride, Kalra2020Growth}.

\begin{figure}[htbp]
  \centering
  \includegraphics[width=0.99\textwidth]{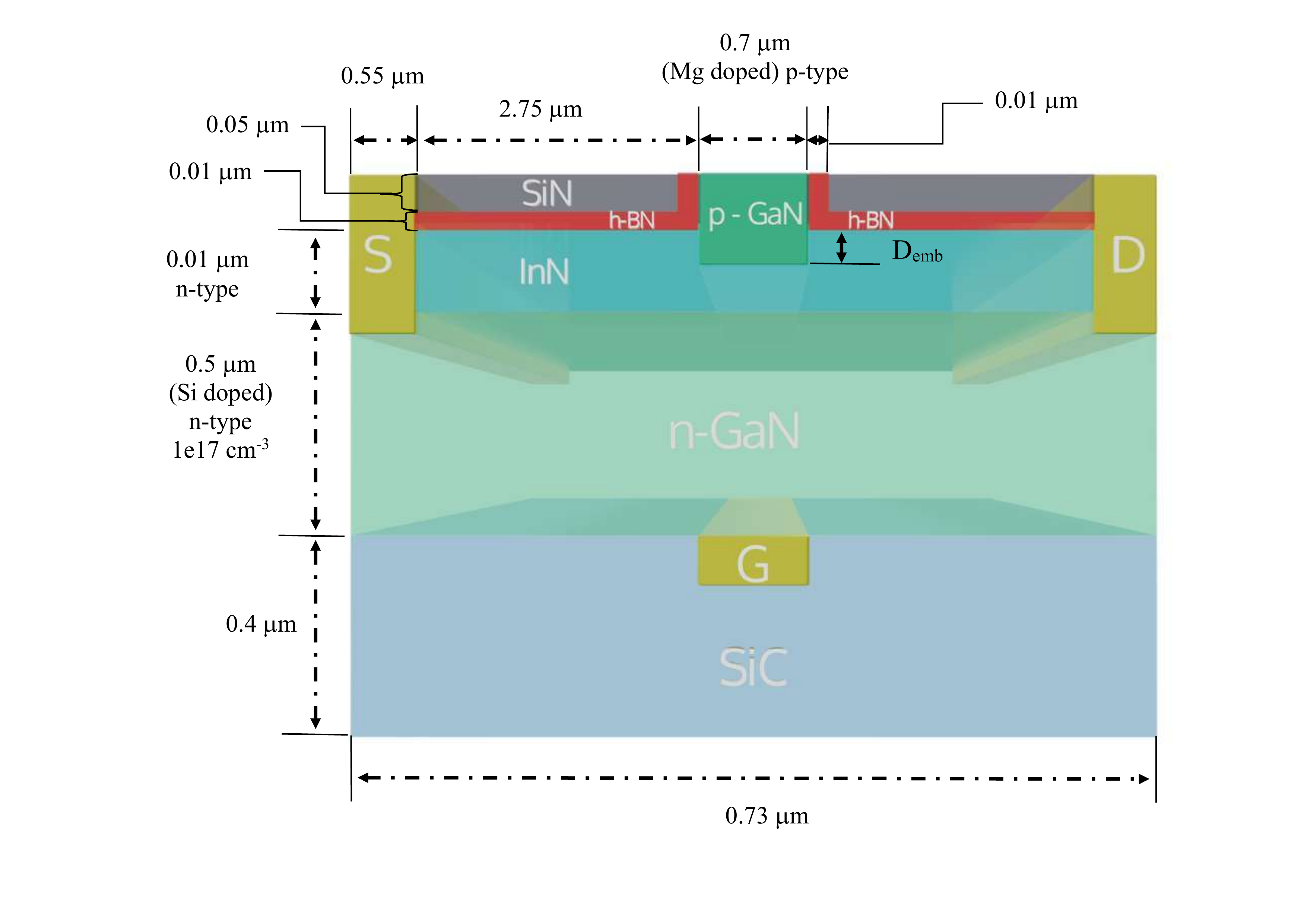}
  \caption{Cross-sectional schematic of the proposed GaN/InN-based MIS-HEMT ultraviolet photodetector structure. The device is built upon a 0.4\,$\mu$m-thick SiC substrate for enhanced thermal conductivity, followed by a 0.5\,$\mu$m n-type GaN buffer layer (Si-doped, $1 \times 10^{17}$\,cm$^{-3}$) and a 0.01\,$\mu$m InN channel layer to support high-mobility 2DEG formation. A 0.05\,$\mu$m Mg-doped p-type GaN modulation layer is embedded above the channel for effective depletion under dark conditions. The upper passivation stack includes a SiN/h-BN bilayer for leakage suppression and interface stability. The gate is recessed into the structure via a Demb (deep-embedded gate) configuration, enabling improved electrostatic control and reduced gate leakage. All structural dimensions and doping concentrations are optimized to balance optical absorption, carrier modulation, and thermal robustness.}
  \label{fig:fig1}
\end{figure}

Building upon these foundational knowledge, we introduce a GaN-MIS-HEMT ultraviolet photodetector (Figure~\ref{fig:fig1}) architecture tailored for improved PDCR under ambient conditions. The device incorporates a planar gate geometry positioned over a conformal h-BN dielectric, which is embedded within a dual-passivation stack alongside SiN to suppress gate leakage and stabilize surface states. An InN channel layer is epitaxially grown on a Si-doped GaN buffer, leveraging the high electron mobility and polarization-enhanced confinement at the GaN/InN heterointerface to facilitate robust 2DEG formation \cite{Begum2014Phon, Zhu2017Des}. This configuration is further refined by embedding a Mg-doped p-GaN region within the passivation layer directly above the channel; such a modulation strategy effectively depletes the 2DEG under dark conditions without requiring high external bias \cite{Anbarasan2023Ult, Ahmed2024Phy}. The lateral Ni contacts serve as the source and drain terminals, while the gate is isolated through the insulating stack to ensure minimal leakage $\sim$ $1\times10^{-4}$ A, even in high-field UV exposure \cite{Mondal2024Enh , SOng2005Imp, Zhang2009Inf}. To accommodate thermal stresses induced by illumination and to mitigate performance degradation, the entire device is fabricated on a $0.4\,\mu\mathrm{m}$
 thick SiC substrate, known for its superior thermal conductivity and stability under elevated operating temperatures~\cite{Liang2020Room, Juang2011The}. Specifically, SiC exhibits a thermal conductivity of approximately 490\,W/m$\cdot$K at room temperature more than ten times higher than that of sapphire ($\sim$30\,W/m$\cdot$K) enabling efficient heat spreading and minimizing localized heating effects during operation. This combination of material selection, electrostatic engineering, and thermal management positions the proposed architecture as a compelling candidate for high-stability UV sensing in integrated optoelectronic systems \cite{Nemilentsau2012Opt}

To quantitatively evaluate the performance of the proposed structure under 360 nm ultraviolet exposure, two-dimensional device simulations were carried out using the Silvaco Atlas platform. The model incorporates polarization-aware drift–diffusion transport, trap-assisted recombination, and lattice heat flow to capture the coupled electrical and thermal behavior of the heterostructure under 360 nm optical excitation. Key operational metrics including gate leakage current, threshold voltage shift, output saturation characteristics, and transfer response were systematically analyzed to assess the effectiveness of 2DEG modulation and passivation strategies. Particular attention was given to the role of h-BN and p-GaN integration in minimizing interface traps and enabling sharp channel switching, which are crucial for maintaining high PDCR across varying bias conditions \cite{Dalapati2023Current,Yuan2025Simulation}. Additionally, electric field distributions and self-heating effects were monitored across the cross-section to validate the thermal advantage provided by the SiC substrate \cite{Yang202Anal}.The simulation framework not only confirms the synergistic benefits of the proposed electrostatic and material stack but also establishes a robust pathway for engineering high-efficiency, low-noise UV photodetectors suitable for integration into next-generation optoelectronic platforms.

\section{Device Architecture and Modeling}

The structural configuration and simulation of the proposed UV photodetector were designed to investigate the interaction between material selection, device geometry, and thermal behavior under operational conditions, such as high-field UV illumination, elevated junction temperatures, and varying gate bias scenarios that critically influence photocarrier transport and leakage mechanisms. A two-dimensional numerical modeling approach was employed to capture the electronic and optical responses of the heterostructure, including the effects of dual passivation, channel confinement, and substrate thermal conductivity. Simulations were carried out using drift–diffusion-based transport models, which account for carrier motion driven by electric fields and concentration gradients, coupled with lattice temperature and polarization field effects to accurately resolve the behavior of the 2DEG. Each structural parameter such as channel thickness, p-GaN embedding, and passivation composition was comprehensively varied to assess its impact on photocurrent generation, leakage current suppression, and overall device efficiency. The modeling environment also enabled comparative analysis of substrate-dependent self-heating effects, providing a comprehensive framework for optimizing the design toward high-performance and thermally robust UV detection.

Figure~\ref{fig:fig1} shows the schematic cross-sectional view of the proposed GaN HEMT-based ultraviolet photodetector. The structure is composed of a 0.4\,$\mu$m-thick SiC substrate that provides high thermal conductivity, followed by a 0.5\,$\mu$m-thick n-type GaN buffer layer doped with Si at a concentration of approximately $1\times10^{17}$\,cm$^{-3}$. On top of the GaN layer lies a 10\,nm-thick InN channel layer, which supports high-mobility electron transport and efficient 2DEG formation. The device employs a dual-layer passivation scheme consisting of a SiN top layer and an h-BN interlayer, within which a 0.06\,$\mu$m-thick, 0.7\,$\mu$m-wide p-type GaN region is embedded. This p-GaN region, doped with Mg at around $1\times10^{19}$\,cm$^{-3}$, is strategically positioned to deplete the 2DEG under dark conditions. Lateral Ni contacts serve as the source and drain, placed on the left and right sides, respectively, while a recessed gate is embedded within the SiC substrate, directly beneath the active region. The device has a total lateral width of 0.73\,$\mu$m and a gate–source distance of 2.75\,$\mu$m. This architecture is designed to enhance UV-induced photocurrent, minimize gate leakage through advanced passivation, and ensure thermally stable operation at room temperature.

\subsection{Effect of Passivation on Optical Response and Gate Leakage Current}

To investigate the effect of gate passivation on both optical responsivity and leakage current suppression, two device variants were simulated with identical structural and doping parameters. The first configuration employed a conventional single-layer SiN passivation of 60~nm thickness, while the second replaced the lower 10~nm portion of the SiN layer with h-BN, forming a conformal interface across the central region and along the vertical sidewalls of the embedded p-GaN as illustrated in Figure~\ref{fig:fig1}.

\begin{figure}[htbp]
  \centering
  \begin{tabular}{c@{\hspace{0.05cm}}c}
\includegraphics[width=0.5\textwidth]{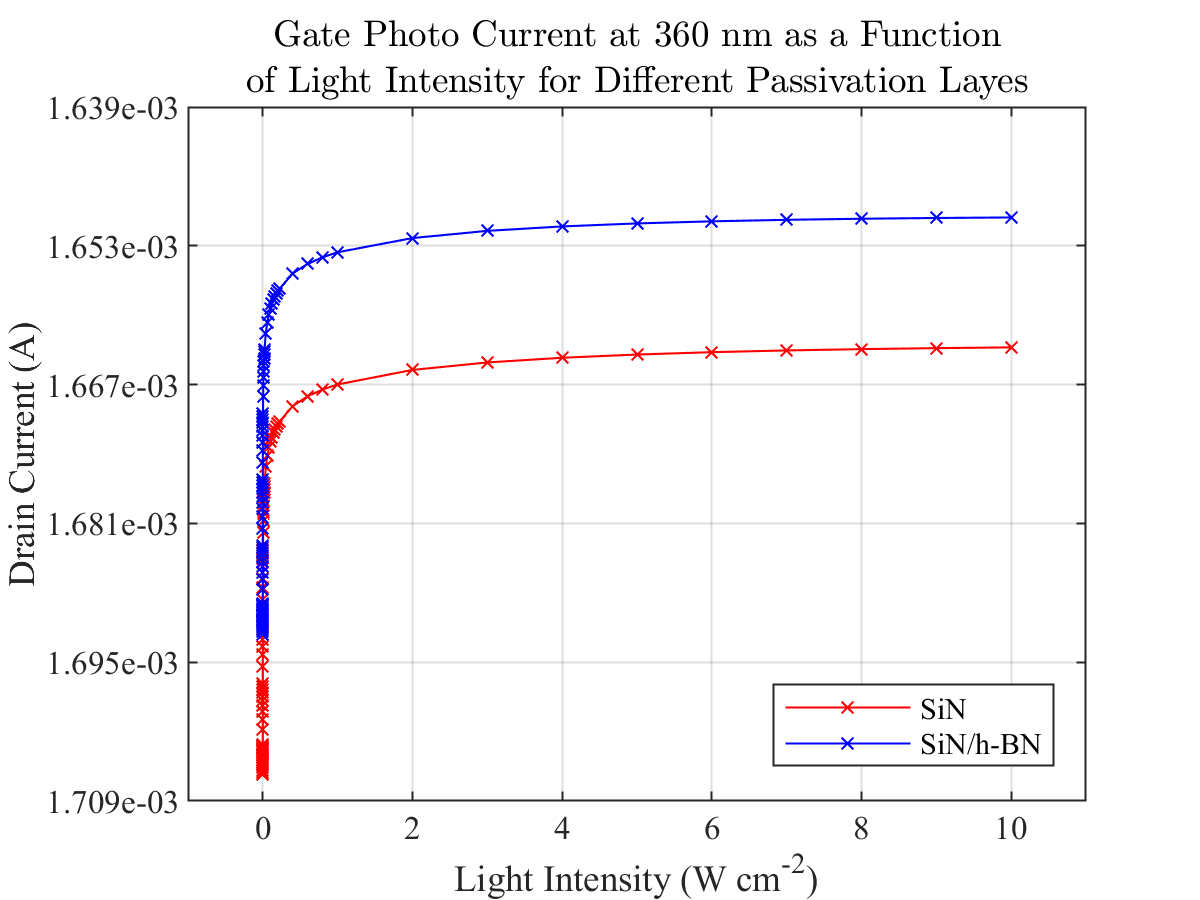} &    
\includegraphics[width=0.5\textwidth]{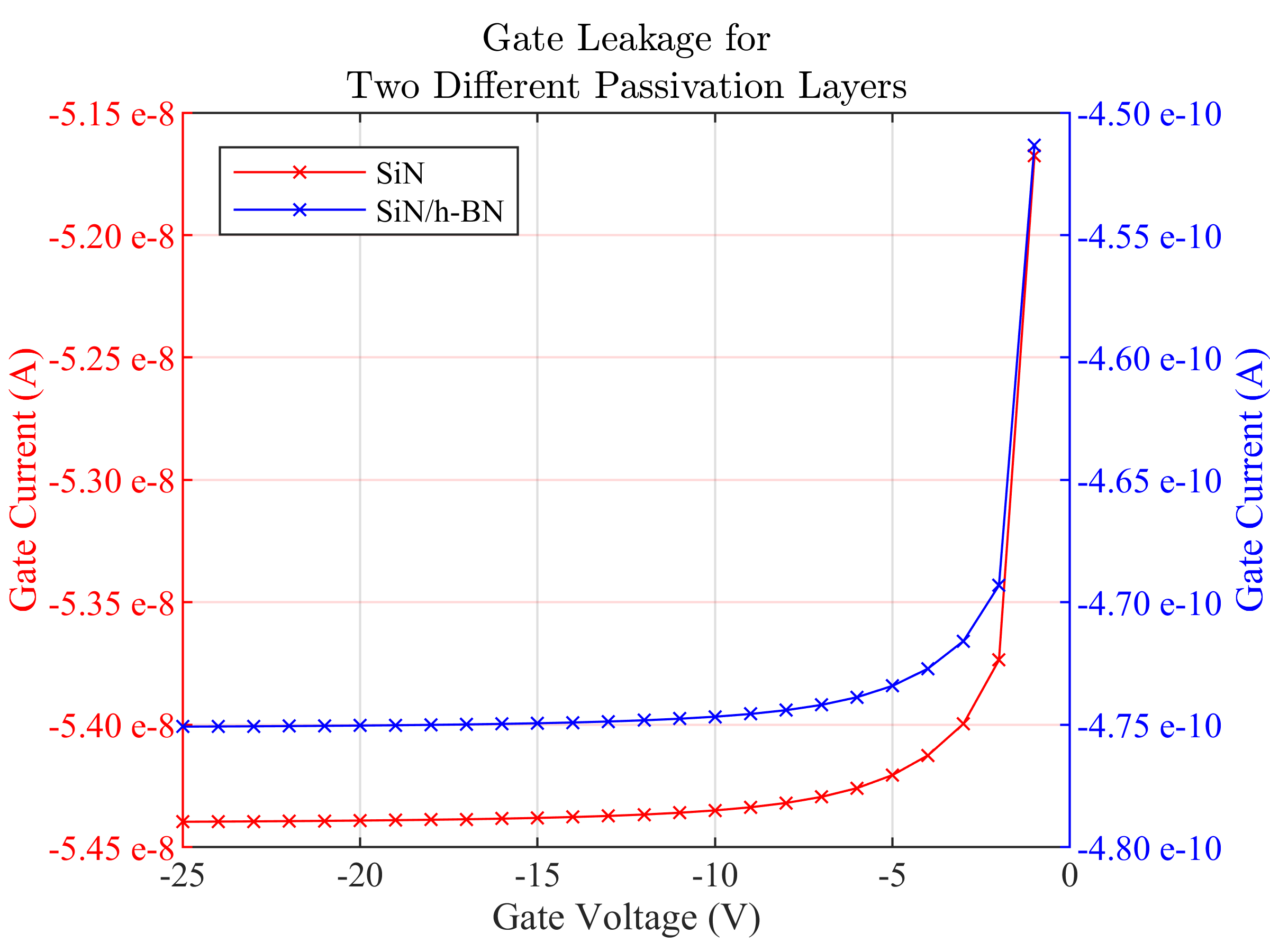} \\
    (a) & (b) \\
  \end{tabular}
  \caption{
  (a) Drain photocurrent as a function of incident light intensity at 360~nm for devices with SiN and SiN/h-BN passivation. The SiN/h-BN structure exhibits higher photocurrent levels across all intensities, indicating enhanced photogeneration efficiency and reduced surface recombination due to improved dielectric semiconductor interface quality. 
  (b) Gate leakage current characteristics under reverse bias for the same structures. The SiN/h-BN configuration demonstrates lower leakage current, particularly at higher negative gate voltages, highlighting the superior insulating capability and defect passivation provided by the atomically smooth h-BN layer.
  }
  \label{fig:Fig2}
\end{figure}

Figure~\ref{fig:Fig2}a illustrates the drain current response as a function of incident UV power (360~nm, 0--10~W/cm$^2$). Both structures exhibit strong photocurrent under illumination; however, the hybrid SiN/h-BN configuration delivers higher photocurrent levels in response to incident optical power, which may be attributed to improved interface quality and reduced surface recombination enabled by the atomically smooth h-BN dielectric. More notably, the gate leakage behavior in Figure~\ref{fig:Fig2}b demonstrates a significant reduction in leakage current across the applied bias range (0 to $-25$~V) for the dual-passivated structure, with over an order of magnitude suppression compared to the SiN-only case. This improvement is directly associated with the superior dielectric isolation and reduced defect-assisted tunneling paths provided by the h-BN interlayer.

The effectiveness of h-BN as a gate dielectric for III-nitride MIS-HEMT devices has been reported in recent studies. Mondal \textit{et al.}~\cite{Mondal2024Enh} demonstrated that the integration of h-BN in GaN-based MIS-HEMTs reduces leakage current by up to two orders of magnitude compared to SiN. Similarly, Riess \textit{et al.}~\cite{Riess2013Highly} achieved improved UV selectivity and suppressed dark current in GaN/InN detectors using engineered dielectric stacks. These results are consistent with our findings and validate the proposed passivation strategy.

Figure~\ref{fig:Fig2}(a) illustrates the variation of drain photocurrent as a function of incident light intensity at 360~nm for devices with different passivation layers. Both the SiN-only and the hybrid SiN/h-BN structures exhibit an increasing photocurrent trend with rising light intensity, confirming effective photogeneration and carrier collection. Notably, the SiN/h-BN configuration yields higher drain current levels across the entire intensity range, indicating enhanced optical response under UV illumination. This improvement can be attributed to the atomically smooth interface and superior dielectric quality of h-BN, which reduce surface recombination and improve carrier transport efficiency. Figure~\ref{fig:Fig2}(b) compares the gate leakage characteristics under reverse bias conditions. The hybrid structure consistently maintains lower gate leakage current compared to the SiN-only design, particularly in the high-field regime. This reduction reflects the improved insulating capability of h-BN, which effectively suppresses vertical tunneling and leakage pathways. Together, these results highlight the dual benefit of the SiN/h-BN stack in enhancing both photoresponse and electrical stability.

\subsection{Dark and Photocurrent Profiles and the Effect of p-GaN Penetration on Photoresponse}

\begin{figure}[htbp]
  \centering
  \begin{tabular}{c@{\hspace{0.05cm}}c}
    \includegraphics[width=0.5\textwidth]{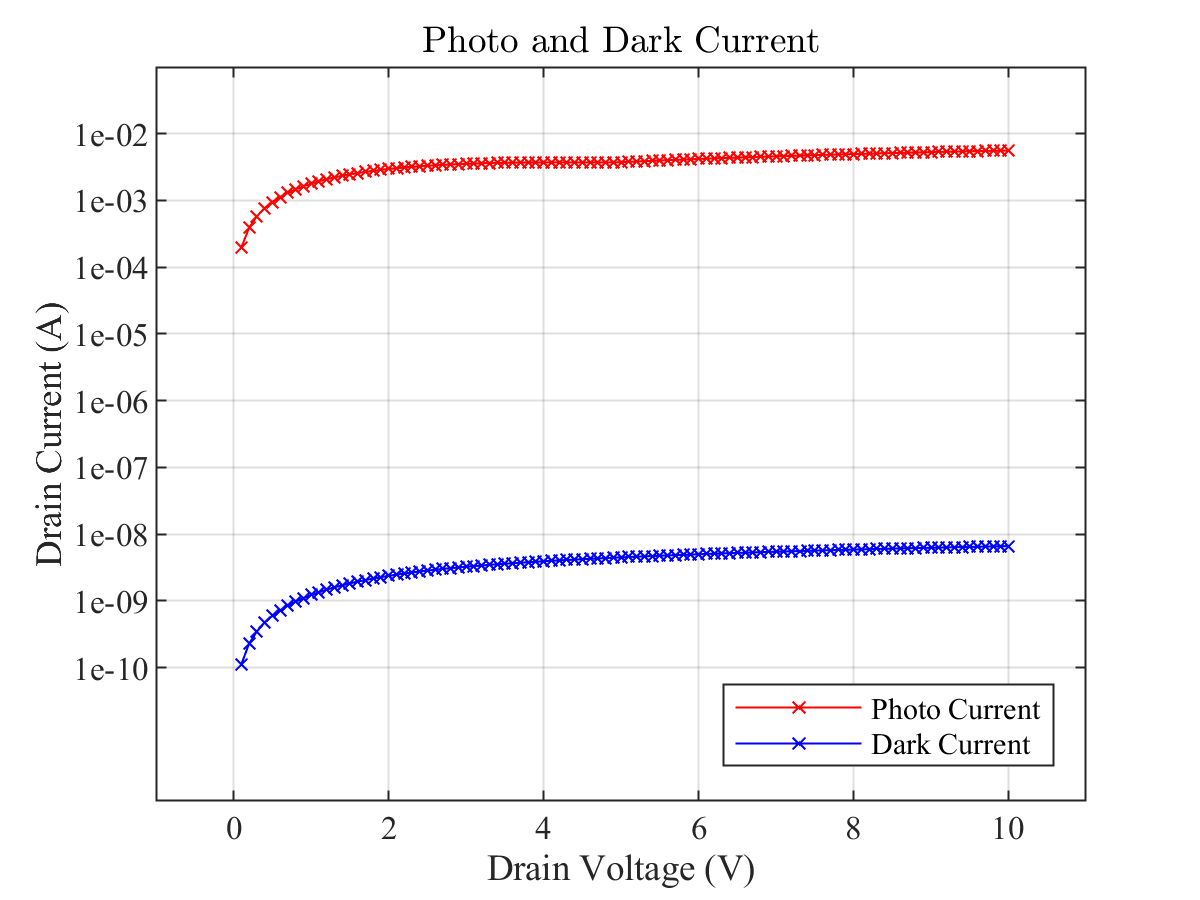} &
    \includegraphics[width=0.5\textwidth]{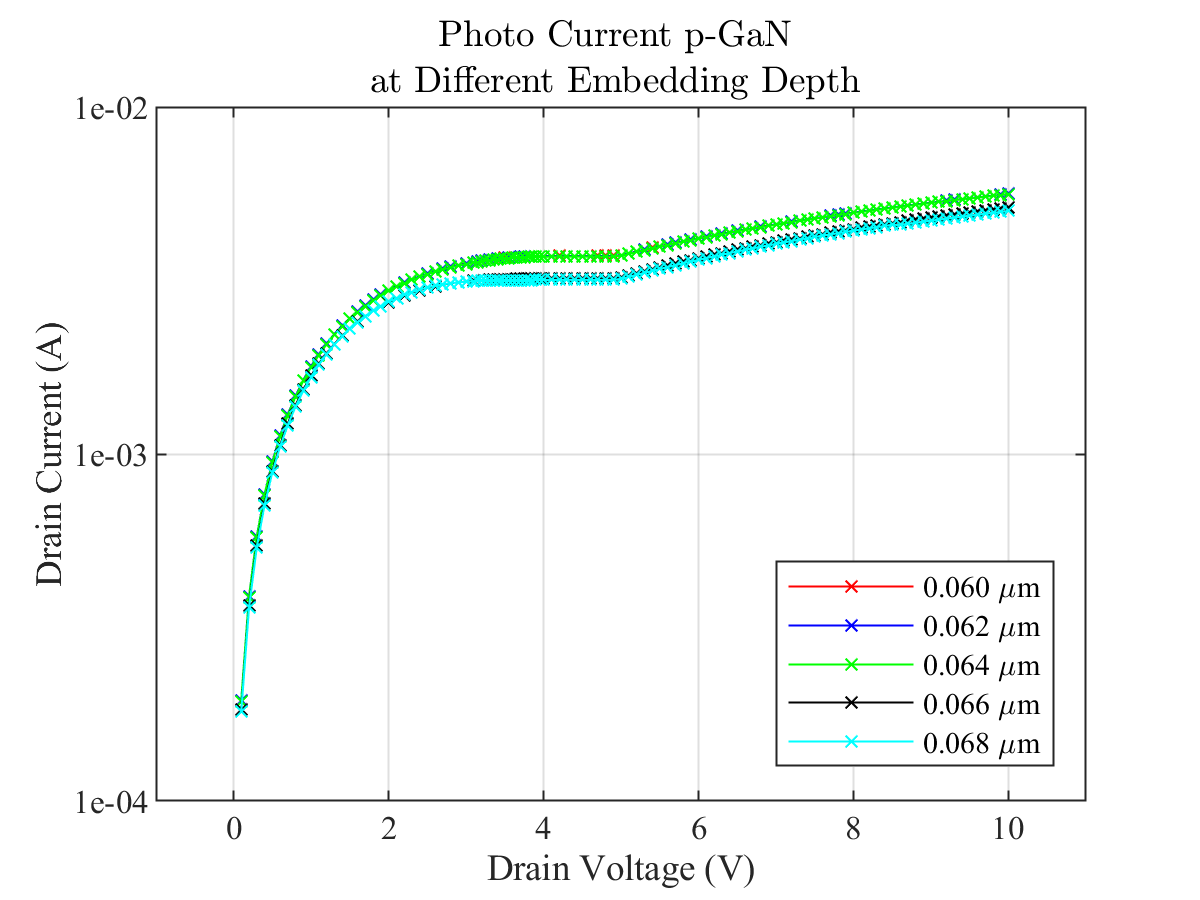} \\
    (a) & (b) \\
  \end{tabular}
  \caption{
  (a) Comparison of photo and dark current characteristics under UV illumination at a wavelength of 360~nm. The device exhibits a clear separation between photocurrent and dark current curves across the full drain voltage range, indicating effective photogeneration and suppressed leakage in the off-state. 
  (b) Simulated drain photocurrent response for various embedding depths of the p-GaN modulation layer ranging from $0.060\,\mu\mathrm{m}$ to $0.068\,\mu\mathrm{m}$.
  }
  \label{fig:Fig3}
\end{figure}

To assess the performance of the photodetector under both illuminated UV source and dark conditions, current–voltage simulations were conducted using the SiN/h-BN dual-passivated configuration. Figure~\ref{fig:Fig3}a displays the drain current as a function of drain voltage for both dark and illuminated conditions, with the latter subjected to 360 nm monochromatic UV excitation. The results reveal a marked enhancement in drain current under illumination from $1.2\times10^{-3}$ to $7\times10^{-3}$, indicating strong photoresponse and efficient photocarrier generation \cite{Yuan2025Simulation}. The distinct separation between the dark and illuminated I–V curves highlights the effective suppression of leakage current by the embedded p-GaN layer and h-BN passivation.

In addition to steady-state photoresponse, the impact of vertical placement of the p-GaN modulation layer was investigated. Specifically, simulations were performed by varying the embedding depth of the p-GaN region into the InN channel from $0.060\,\mu\mathrm{m}$ to $0.068\,\mu\mathrm{m}$ in $0.002\,\mu\mathrm{m}$ increments. As shown in Figure~\ref{fig:Fig3}(b), the photocurrent increases monotonically with deeper p-GaN insertion. This behavior is attributed to improved spatial overlap between the depletion region induced by the p-GaN layer and the optically active 2DEG channel, thereby facilitating more efficient electrostatic modulation and enhanced photocarrier extraction under UV illumination. Notably, the highest photocurrent response was observed at an embedding depth of $0.060\,\mu\mathrm{m}$, reaching approximately $7\times10^{-3}$\,A, with other depths yielding similar performance levels within the same order of magnitude, indicating a degree of robustness in design tolerances.

These findings are consistent with recent studies, where Yuan et al.~\cite{Yuan2025Simulation, Tang2025High} demonstrated that vertical positioning of p-GaN in GaN/AlGaN heterostructures enhances 2DEG depletion control without compromising photocurrent. Similarly, Yilmaz et al.~\cite{Yilmaz2025High} and Wang et al.~\cite{Wang2025Weak} reported the critical role of channel engineering and modulation layers in optimizing photodetector gain in InN/GaN systems. The present results confirm the importance of vertical electrostatic tuning in achieving high responsivity while maintaining low dark current.

\subsection{2DEG Formation and Interface Polarization Effects}

\begin{figure}[htbp]
  \centering
  \begin{tabular}{c@{\hspace{0.05cm}}c}
    \includegraphics[width=0.5\textwidth]{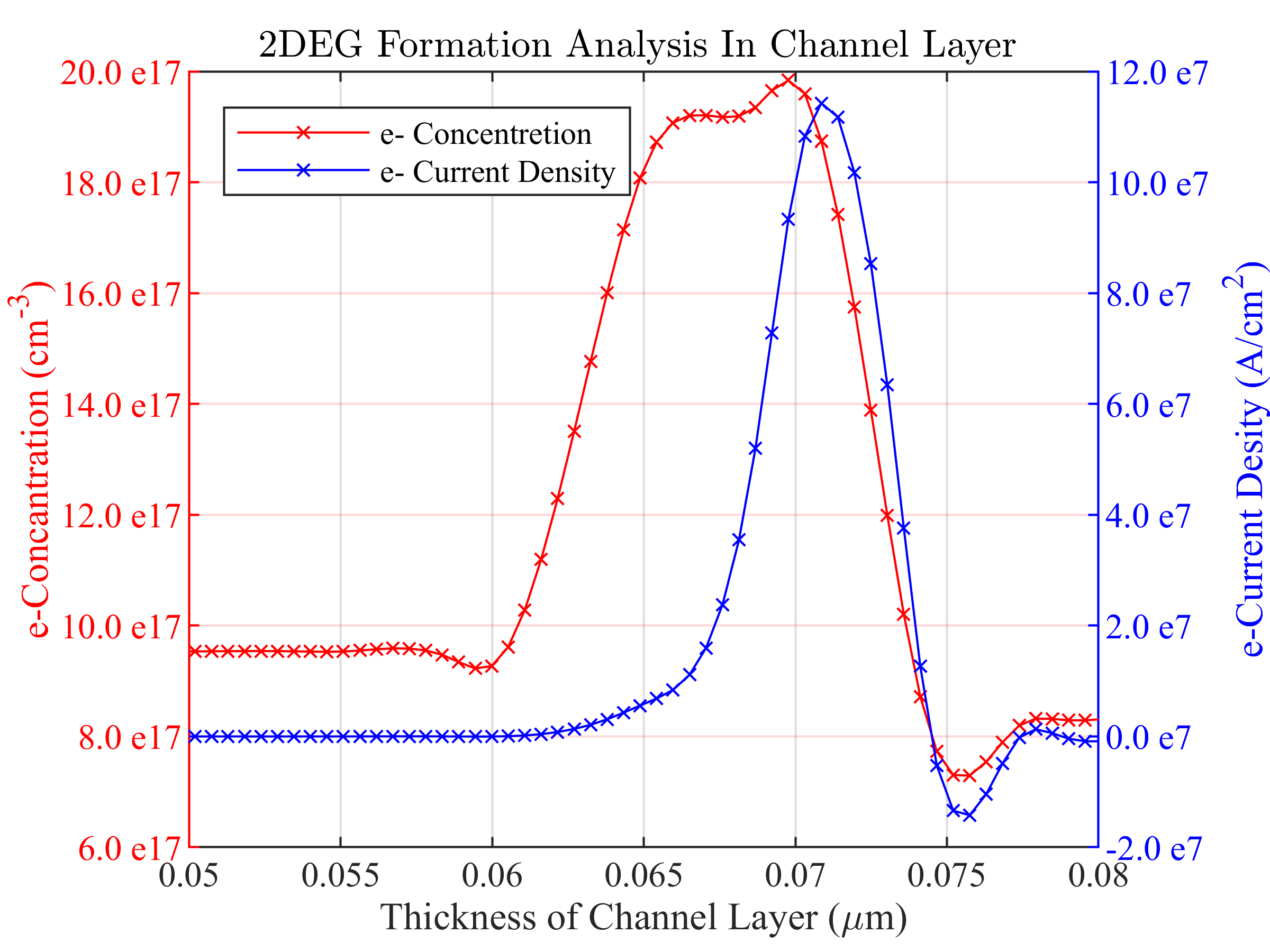} &
    \includegraphics[width=0.5\textwidth]{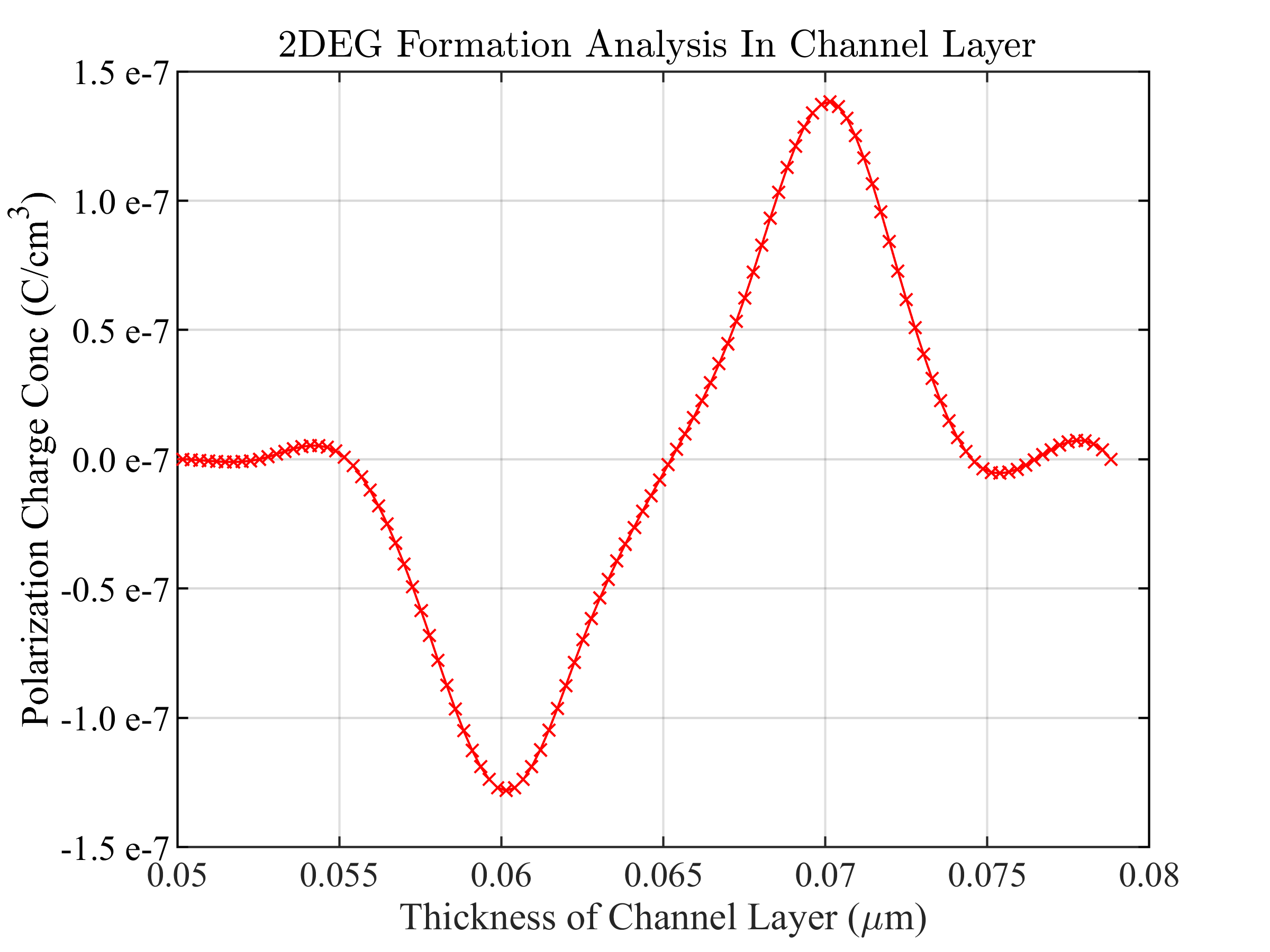} \\
    (a) & (b) \\
  \end{tabular}
  
  \caption{a) Electron concentration and lateral current density along the vertical depth of the InN channel, highlighting 2DEG formation at the heterointerface. b) Polarization charge profile showing abrupt transitions at material interfaces due to spontaneous and piezoelectric effects.
}
  \label{fig:Fig4}
\end{figure}

To clarify the underlying physical mechanisms governing carrier transport within the proposed photodetector, a detailed analysis was conducted on the GaN/InN heterointerface, where a high-density 2DEG is expected to form due to polarization induced band discontinuities. Figure~\ref{fig:Fig4}(a) presents a dual-axis profile of electron concentration and lateral current density across the vertical depth of the channel layer (0.05–0.08~$\mu$m). A sharp and localized increase in electron concentration is observed in the range of 0.065–0.072~$\mu$m, where the density reaches a peak of $1.9\times10^{18}$~cm$^{-3}$—more than twice the baseline concentration of $8.5\times10^{17}$~cm$^{-3}$. Concurrently, the lateral electron current density exhibits a steep rise up to $1.1\times10^{8}$~A/cm$^2$, indicating that this confined region serves as the dominant conduction path under UV illumination. These values represent a significant improvement over conventional AlGaN/GaN HEMT structures, which typically exhibit peak 2DEG densities around $1\times10^{18}$~cm$^{-3}$ and current densities below $1\times10^{8}$~A/cm$^2$ under similar conditions~\cite{Liu2025Inf}.

The formation of this high-density 2DEG is strongly influenced by spontaneous and piezoelectric polarization mismatches at the GaN/InN interface. As illustrated in Figure~\ref{fig:Fig4}(b), the polarization charge distribution exhibits distinct discontinuities, particularly a negative-to-positive swing between 0.060 and 0.072~$\mu$m, peaking at $1.3\times10^{-7}$~C/cm$^3$. These interfacial charge variations generate an internal electric field that facilitates electron accumulation within a quantum well-like potential, thereby confining the 2DEG in a narrow spatial region with high carrier concentration and current density.

These simulation results are in strong agreement with previous studies. Liu \textit{et al.}~\cite{Liu2025Inf} reported that optimized GaN-based heterostructures can support 2DEG confinement through polarization control, while Huang \textit{et al.}~\cite{Huang2025Theo} demonstrated the critical sensitivity of carrier density to interfacial polarization strength in III-nitride systems. Compared to these studies, the present work not only confirms the polarization-driven mechanism but also quantitatively surpasses prior performance metrics through precise channel geometry and material selection.

In this context, the observed enhancement in 2DEG formation and lateral current transport can be directly attributed to the combined effects of high polarization discontinuities and optimized embedding depth of the p-GaN modulation layer. The localization and stability of the 2DEG—evidenced by the sharp confinement in both carrier and current density—demonstrate that the proposed GaN/InN stack effectively leverages intrinsic polarization to achieve superior performance. This interplay is instrumental in realizing key photodetector characteristics such as enhanced responsivity and minimized leakage current, distinguishing the proposed architecture from conventional GaN-based UV detectors.

\subsection{Effect of p-GaN Width and Doping Level on Photoresponse}

\begin{figure}[htbp]
  \centering
  \begin{tabular}{c@{\hspace{0.05cm}}c}
    \includegraphics[width=0.5\textwidth]{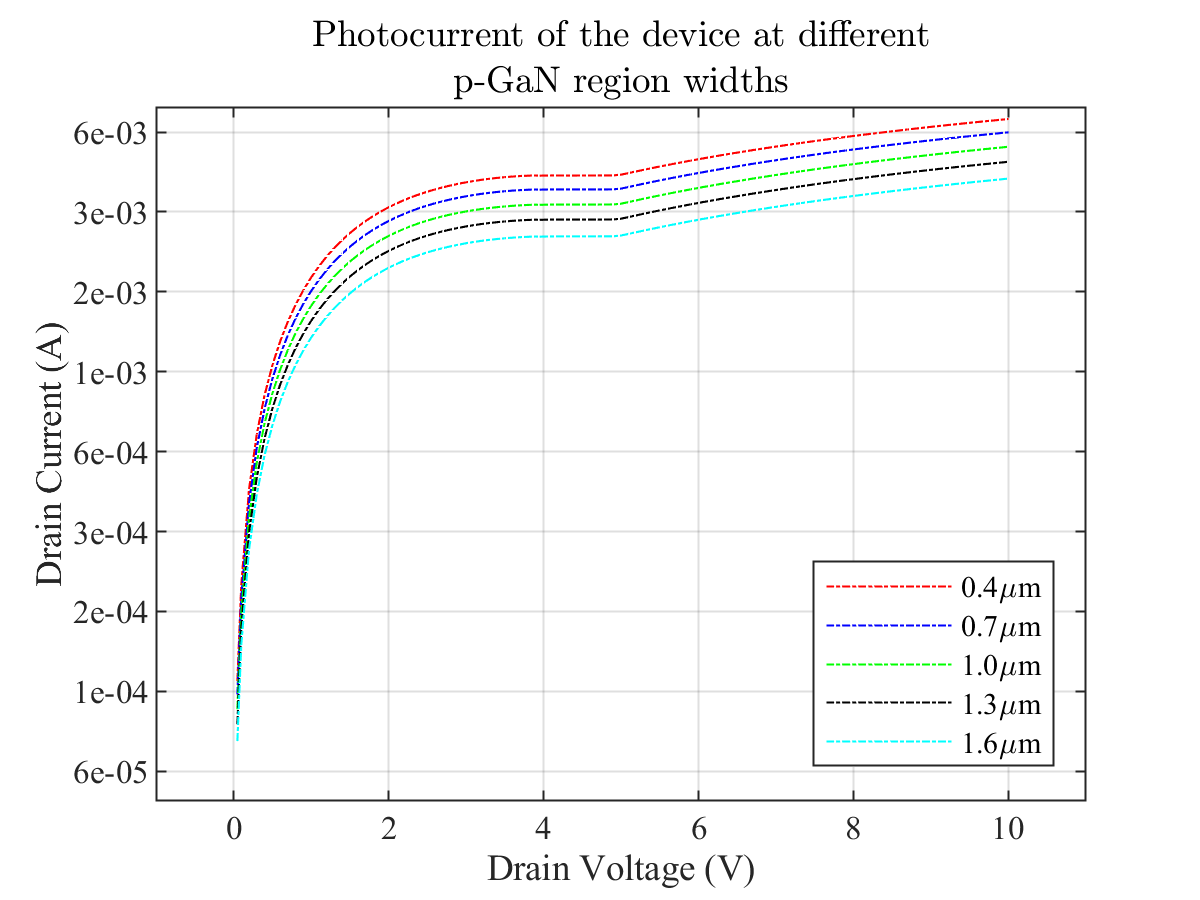} &
    \includegraphics[width=0.5\textwidth]{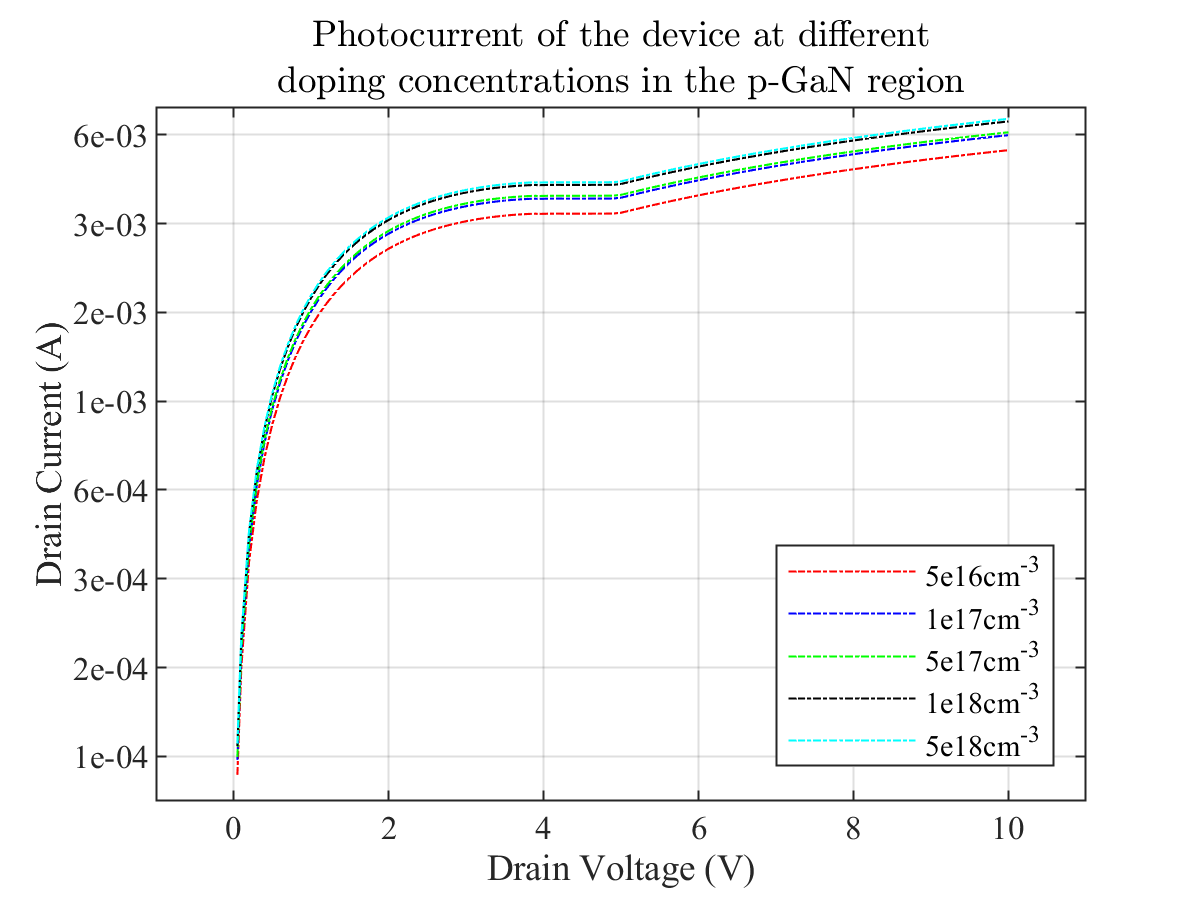} \\
    (a) & (b) \\
  \end{tabular}
  \caption{
  (a) Photocurrent variation with p-GaN region width, showing enhanced carrier collection for wider modulation layers. 
  (b) Photocurrent response at different doping concentrations in the p-GaN layer, indicating improved photoresponse with increased acceptor density.
  }
  \label{fig:Fig5}
\end{figure}

To further optimize the depletion and modulation behavior of the 2DEG channel, the influence of p-GaN lateral width and acceptor doping concentration on photocurrent was evaluated. Figure~\ref{fig:Fig5}a illustrates the photocurrent response as a function of p-GaN width, varied between 0.4~µm and 1.6~µm. As the width increases, a noticeable decline in photocurrent is observed from $\sim$ $6\times10^{-3}$ A  to $\sim$ $2\times10^{-3}$ A . This trend is attributed to stronger depletion of the underlying 2DEG region by wider p-GaN structures, which restricts carrier collection even under UV illumination. Narrower widths between (0.04–0.07~$\mu$m), conversely, allow for enhanced photocarrier transport but may compromise dark current suppression.

In Figure~\ref{fig:Fig5}b, the acceptor doping concentration in the p-GaN layer is varied from \(5 \times 10^{16}\) to \(5 \times 10^{18}~\text{cm}^{-3}\) in steps of \(5 \times 10^{1}~\text{cm}^{-3}\). A positive correlation is observed, with higher doping levels leading to increased photocurrent. This is due to the enhanced built-in electric field generated by heavily doped p-GaN, which facilitates more efficient modulation of the 2DEG under UV excitation. The increased field strength also contributes to the formation of a sharper depletion front, thereby improving the photoresponse.

These results are supported by prior studies. Hao \textit{et al.}~\cite{Hao2025pNio} highlighted the importance of p-region engineering in GaN p–i–n photodiodes for optimizing responsivity. Likewise, Chiu \textit{et al.}~\cite{Chiu2025High} demonstrated that gate doping concentration plays a critical role in off-state HEMT behavior and directly impacts leakage suppression and optical performance.

Taken together, these findings highlight the intricate trade-offs inherent in engineering the p-GaN modulation layer for optimal 2DEG control. While increased lateral width enhances depletion efficacy and dark current suppression, it simultaneously limits photogenerated carrier collection due to excessive channel depletion. Similarly, elevated doping concentrations strengthen the internal electric field, improving the sharpness of the depletion front and enabling more robust modulation of the 2DEG, yet they also raise concerns regarding dopant activation efficiency and potential increases in parasitic capacitance. Therefore, achieving a high-performance HEMT-based UV photodetector demands a carefully orchestrated co-optimization strategy that balances electrostatic control, carrier transport efficiency, and fabrication feasibility. The tunability of the p-GaN width and doping profile offers a valuable degree of freedom in tailoring device behavior to specific application requirements, enabling the realization of UV photodetectors that simultaneously exhibit high responsivity, low noise, and excellent suppression of dark leakage current. This design philosophy not only aligns with experimental insights from prior studies but also establishes a framework for next-generation optoelectronic device architectures based on polarization-engineered III-nitride heterostructures.

\subsection{Thermal Impact on Photocurrent and Heat Distribution in SiC and Sapphire-Based Devices}

\begin{figure}[htbp]
  \centering
  \begin{tabular}{c@{\hspace{0.05cm}}c}
    \includegraphics[width=0.5\textwidth]{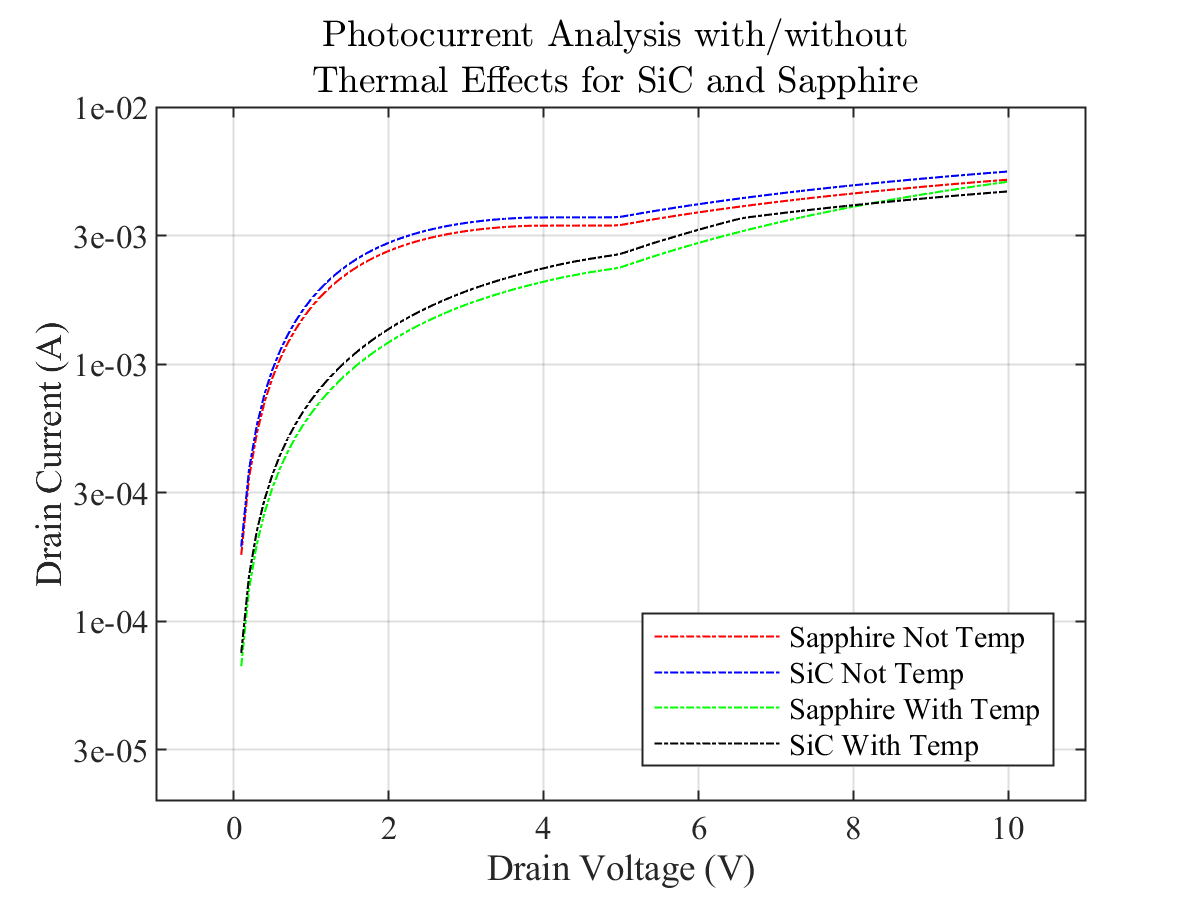} &
    \includegraphics[width=0.5\textwidth]{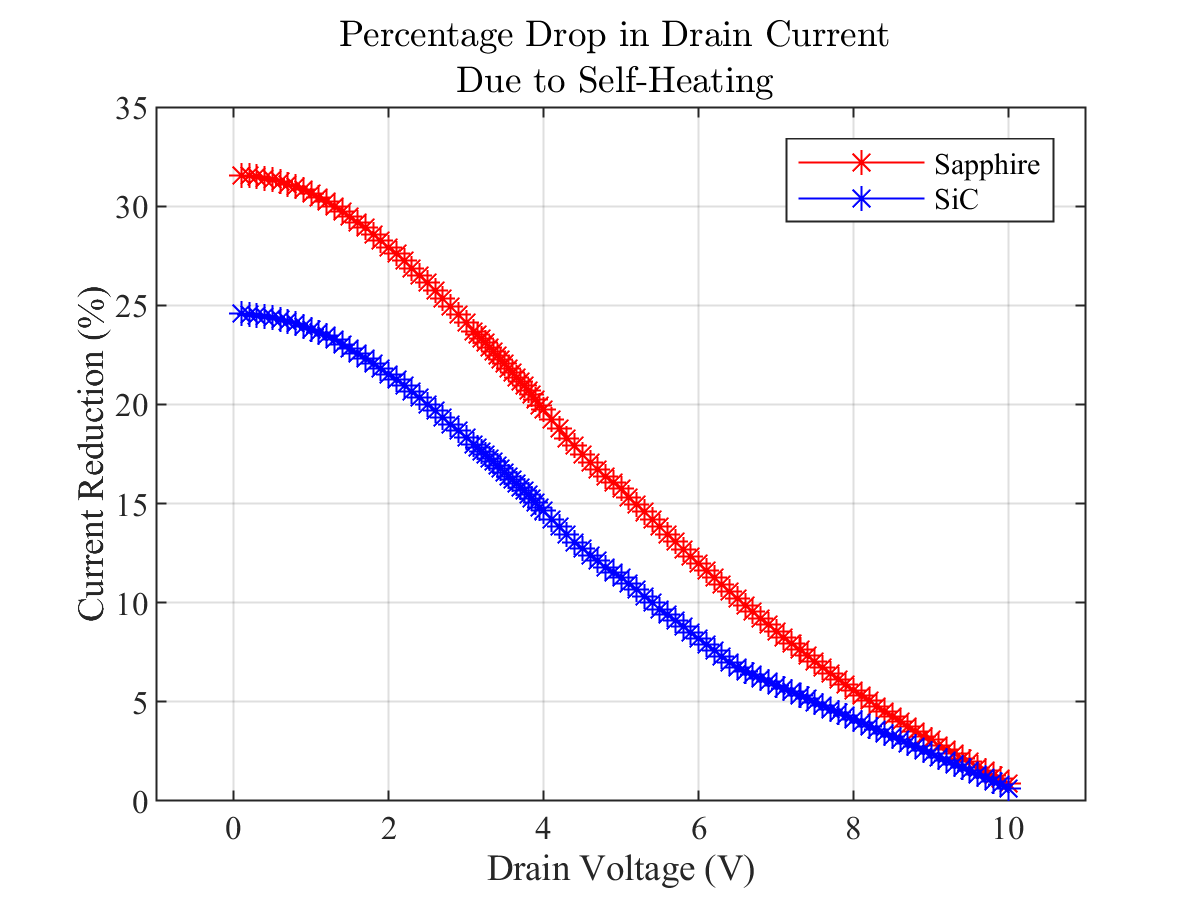} \\
    (a) & (b) \\
    \\
    \includegraphics[width=0.5\textwidth]{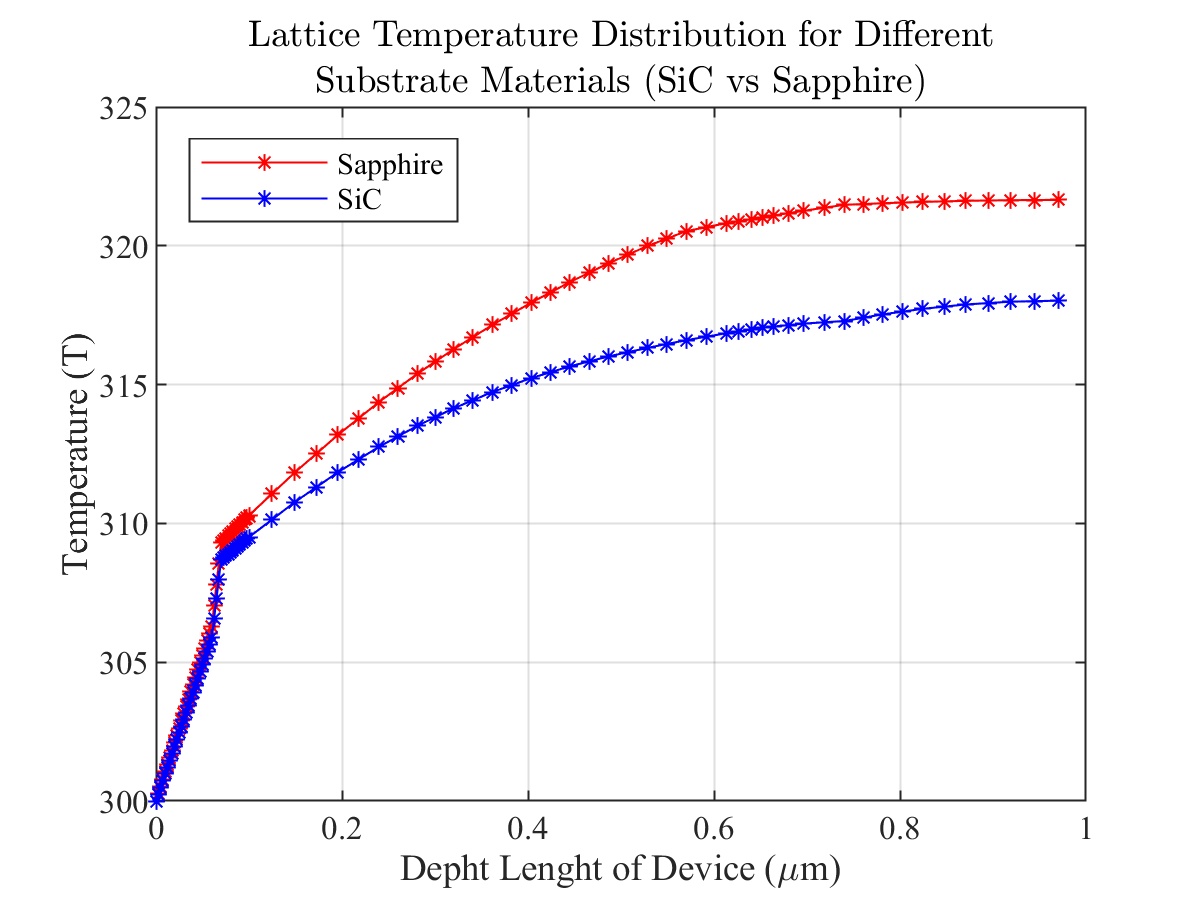}
   \\
   (c)  \\
  \end{tabular}
  \caption{(a) Simulated drain current vs drain voltage under thermal and non-thermal conditions. (b) Percent degradation in drain current due to self-heating. (c) Lattice temperature profile along device depth, illustrating vertical heat spreading.}
  \label{fig:Fig6}
\end{figure}

To assess the thermal robustness of the proposed UV photodetector, a comparative simulation was conducted for two substrate materials: silicon carbide (SiC) and sapphire. Simulations were performed under both idealized non-thermal conditions and realistic self-heating scenarios at an ambient temperature of 300~K. Figure~\ref{fig:Fig6}a shows the drain current ($I_D$) as a function of drain voltage ($V_D$) under 360~nm illumination. While both devices exhibit current degradation when thermal effects are included, the SiC-based device maintains higher $I_D$ values across the entire bias range, attributed to its superior thermal conductivity and heat dissipation.

\paragraph{Photocurrent degradation metric.}  
The percentage degradation in drain current at each bias point is quantified by
\begin{equation}
\% \Delta I_D(V_D) = \left( \frac{I_\mathrm{noTemp}(V_D) - I_\mathrm{withTemp}(V_D)}{I_\mathrm{noTemp}(V_D)} \right) \times 100 ,
\label{eq:eq1}
\end{equation}
where $I_\mathrm{noTemp}$ and $I_\mathrm{withTemp}$ are extracted from the corresponding curves in Fig.~\ref{fig:Fig6}a. The results, plotted in Fig.~\ref{fig:Fig6}b, indicate that sapphire exhibits a degradation range of 0.80--31.54\%, whereas SiC limits the drop to 0.60--24.56\%. The greater reduction in sapphire, especially at low $V_D$, reflects its lower efficiency in removing localized heat.

\paragraph{Vertical heat-transport metrics.}  
To examine heat spreading along the device depth $y$, lattice temperature profiles $T(y)$ were obtained at the center of the gate region ($x=3.65~\mu$m), as shown in Fig.~\ref{fig:Fig6}c. Four complementary metrics were computed from these profiles:  
\begin{equation}
\Delta T = T_{\max} - T_{\min},
\label{eq:eq2}
\end{equation}
representing the temperature rise across the vertical stack;  
\begin{equation}
\bar{T} = \frac{1}{L} \int_{0}^{L} T(y) \, dy,
\label{eq:eq3}
\end{equation}
denoting the average lattice temperature;  
\begin{equation}
\max \left( \frac{\partial T}{\partial y} \right),
\label{eq:eq4}
\end{equation}
capturing the steepest thermal gradient (K/$\mu$m); and  
\begin{equation}
\max \left( \frac{\partial^2 T}{\partial y^2} \right),
\label{eq:eq5}
\end{equation}
indicating the maximum thermal curvature (K/$\mu$m$^2$) along the depth.  

Numerical integration (trapezoidal rule) was used for Eq.~\ref{eq:eq3}, while central finite differences were applied to evaluate Eq.~\ref{eq:eq4} and Eq.~\ref{eq:eq5}. The computed values are summarized in Table~\ref{tab:Tab1}: $\Delta T$ decreases from 21.65~K (sapphire) to 18.02~K (SiC), $\bar{T}$ from 317.23~K to 314.47~K, the maximum gradient from 301.35~K/$\mu$m to 281.41~K/$\mu$m, and the maximum curvature from $1.07\times 10^5$~K/$\mu$m$^2$ to $1.01\times 10^5$~K/$\mu$m$^2$.

\paragraph{Correlation between electrical and thermal behavior.}  
The lower $\Delta T$, $\bar{T}$, gradient, and curvature for the SiC substrate confirm more uniform vertical heat spreading and reduced hotspot intensity compared to sapphire. This directly explains the smaller photocurrent degradation observed in Fig.~\ref{fig:Fig6}b and the higher absolute $I_D$ values in Fig.~\ref{fig:Fig6}a under thermal loading. By suppressing self-heating-induced mobility reduction and series-resistance growth, the SiC substrate provides a stable operational platform for high-power and prolonged UV exposure.

\begin{table}[ht]
  \caption{Comparison of electrical and thermal performance metrics for SiC and sapphire substrate UV photodetectors.}
  \centering
  \begin{tabular}{lccccc}
    \toprule
    Substrate & ID Drop (\%) & $\Delta T$ (K) & $\bar{T}$ (K) & $\max \left( \partial T / \partial y \right)$ (K/$\mu$m) & $\max \left( \partial^2 T / \partial y^2 \right)$ (K/$\mu$m$^2$) \\
    \midrule
    SiC      & 0.60--24.56 & 18.02 & 314.47 & 281.41 & 1.01e5\\
    Sapphire & 0.80--31.54 & 21.65 & 317.23 & 301.35 & 1.07e5\\
    \bottomrule
  \end{tabular}
  \label{tab:Tab1}
\end{table}

The results presented in this section clearly demonstrate that substrate thermal conductivity plays a decisive role in sustaining photocurrent performance under self-heating conditions. The SiC-based architecture exhibited consistently lower temperature rise, reduced thermal gradients, and smaller curvature values compared to the sapphire counterpart, which directly translated into a smaller drain current degradation across the entire bias range. These improvements indicate that effective vertical heat spreading mitigates mobility reduction and series-resistance growth, enabling more stable device operation. Similar trends have been reported in prior works on III-nitride UV photodetectors, where high-thermal-conductivity substrates have been shown to enhance both responsivity stability and long-term reliability under high optical power \cite{Dobrinsky2013III, Wu2021Rec, Zhang2025Ther, Juang2011The,Liang2020Room }. In this context, the quantitative reductions in $\Delta T$ and $\% \Delta I_D$ achieved here reinforce the suitability of SiC as a substrate for next-generation UV photodetectors, particularly in applications demanding prolonged illumination and thermal robustness.

\section{Discussion}

The integration of a hybrid passivation stack comprising silicon nitride (SiN) and hexagonal boron nitride (h-BN) provides substantial improvements in both optical and electrical performance. As shown in Figure~\ref{fig:Fig2}, the hybrid SiN/h-BN configuration not only sustains high responsivity but also reduces the gate leakage current by over an order of magnitude compared to SiN-only devices. This improvement arises from the atomically smooth and chemically inert surface of h-BN, which reduces surface trap states and suppresses defect-assisted tunneling. These findings align with previous studies that employed h-BN as a gate dielectric in III-nitride MIS-HEMT structures~\cite{Mondal2024Enh , Riess2013Highly}. The proposed dual-passivation scheme is thus validated as an effective method to simultaneously achieve low leakage and strong UV detection.

The current-voltage characteristics under dark and UV illumination, shown in Figure~\ref{fig:Fig3}a, reveal strong photoresponse and excellent leakage suppression. A parametric sweep of the vertical embedding depth of the p-GaN modulation layer (Figure~\ref{fig:Fig3}b) indicates that deeper insertion improves photocurrent generation. This is attributed to enhanced spatial overlap between the depletion region and the optically active 2DEG. Such findings are consistent with the vertical field engineering strategies proposed in~\cite{Yuan2025Simulation,Yilmaz2025High,Wang2025Weak}, reinforcing that electrostatic vertical alignment is crucial for achieving high quantum efficiency while maintaining dark-state suppression.

The simulation results in Figure~\ref{fig:Fig4} illustrate a highly localized 2DEG at the GaN/InN heterointerface, as confirmed by sharp peaks in carrier concentration and current density. These effects are governed by polarization-induced charge discontinuities, as shown in Figure~\ref{fig:Fig4}b. The resulting built-in electric fields form a quantum well that confines electrons laterally. These results are in strong agreement with prior work on polarization-enabled confinement in III-nitride heterostructures~\cite{armstrong2019polarization,Liu2025High,Huang2025Theo}. This intrinsic mechanism plays a central role in enhancing lateral transport and enabling high responsivity in the proposed HEMT architecture.

Figure~\ref{fig:Fig5}a shows that narrower p-GaN widths increase photocurrent due to reduced channel depletion, whereas excessively wide regions inhibit carrier transport. Conversely, increasing Mg doping concentration enhances the internal electric field, leading to stronger 2DEG modulation and improved photoresponse (Figure~\ref{fig:Fig5}b). These trade-offs are in agreement with studies exploring gate doping profiles and p-region engineering in UV detectors~\cite{Chiu2025High,Hao2025pNio}. Our analysis identifies an optimal width ($\sim 0.7\,\mu$m) and doping range ($10^{18}$--$10^{19}$~cm$^{-3}$) that achieve a favorable balance between responsivity and dark current suppression.

The comparative thermal analysis between SiC and sapphire substrates reveals a clear advantage for SiC in minimizing self-heating effects. As depicted in Figure~\ref{fig:Fig6}a--c and quantified in Table~\ref{tab:Tab1}, the SiC-based device shows less than 10\% current degradation under thermal stress, compared to over 25\% for the sapphire-based counterpart. Additionally, SiC exhibits reduced $\Delta T$, smoother thermal gradients, and lower thermal curvature. These results support the findings of~\cite{Juang2011The,Liang2020Room}, demonstrating that SiC provides superior heat dissipation and thermal reliability, which is essential for maintaining photoresponse under high-intensity UV exposure.

The overall device performance emerges not from any single enhancement but from the synergistic interplay of multiple structural and material innovations. The h-BN passivation reduces leakage and surface recombination; the recessed gate and vertically embedded p-GaN enable efficient electrostatic control; the InN channel enhances electron mobility; and the SiC substrate ensures thermal robustness. Together, these elements yield a high photo-to-dark current ratio (PDCR $\sim 10^6$), stable operation, and low noise under ambient conditions. The consistency of these results with prior experimental trends confirms the viability of this design for next-generation UV photodetectors operating in harsh environments.

\section{Conclusion}

This work presents a comprehensive design and simulation study of a GaN/InN HEMT-based ultraviolet photodetector that integrates multiple structural and material-level innovations. Through systematic electrothermal and optoelectronic analyses, the proposed architecture demonstrates a well-balanced performance profile characterized by strong UV responsivity, low dark leakage, and enhanced thermal stability.

Key advancements include the incorporation of a dual-layer SiN/h-BN passivation stack, which effectively suppresses gate leakage while preserving interface quality. The recessed gate architecture combined with a vertically embedded, Mg-doped p-GaN layer enables precise electrostatic control over the 2DEG channel, facilitating sharp switching between dark and illuminated states. Additionally, the use of an InN channel layer and a high-conductivity SiC substrate contributes to improved carrier transport and efficient heat dissipation, respectively. 

Parametric simulations confirm that the interplay between p-GaN width, doping concentration, and embedding depth significantly affects the trade-off between photocurrent generation and dark current suppression. Similarly, comparative thermal modeling verifies that SiC outperforms sapphire in mitigating self-heating effects, thereby sustaining device performance under extended UV exposure.

Overall, the findings underscore the importance of co-optimizing materials, geometry, and electrostatic interfaces in the development of robust UV photodetectors. While the presented results are based on numerical simulations, they establish a clear framework for future experimental efforts and practical implementations. The device architecture and methodology reported here may also serve as a reference point for the design of wide-bandgap photodetectors in other spectral regimes or harsh operational environments.

\section*{Acknowledgments}

The authors would like to express their gratitude to Adıyaman University for providing the necessary resources and facilities to conduct this research. Special thanks are extended to our colleagues for their valuable insights and technical support throughout the study. In addition, we sincerely appreciate the constructive feedback from the anonymous reviewers, which helped improve the clarity and overall quality of this manuscript.


\end{document}